# Electrically-Controlled Hybrid Superconductor-Ferromagnet Cell for High Density Cryogenic Memory

I. P. Nevirkovets[1] and O. A. Mukhanov[2]

[1] Department of Physics and Astronomy, Northwestern University, 2145 Sheridan Road, Evanston, IL 60201, USA

[2] SEEQC, Inc., 150 Clearbrook Road, Elmsford, NY 10523, USA

**Abstract**

We report the fabrication and testing, at 4.2 K, of an $S_1IS_2FS_3$ device, where S, F, and I denote a superconductor (Nb), a ferromagnetic material (permalloy), and an insulator ($AlO_x$), respectively. The F layer covers about one half of the top electrode of the $S_1IS_2$ Josephson junction and is positioned off-center. Electric current, $I_{tr}$, along the $S_3$ electrode can change the magnetization of the F layer in such a way that, for one direction of $I_{tr}$, a magnetic flux penetrates the junction perpendicular to the layers, whereas for the opposite direction, the perpendicular magnetic flux can be removed. In the former state, the modulation pattern of the Josephson critical current, $I_c$, in the magnetic field, $H$, may acquire minimum near $H=0$, and restores its usual shape with maximum in the second state. These states can be used for building a compact cryogenic memory compatible with single flux quantum electronics.

In recent years, a significant research effort has been dedicated for development of cryogenic memory for use in the single flux quantum (SFQ) circuits [1,2]. Under the recent IARPA C3 program [3,4], several potential memory technologies were investigated [5-13]. Other proposals for cryogenic memory include traditional SQUID-based technology [14], adaptation of CMOS memory [15], SISFS junctions (here S, I, and F denote a superconductor, an insulator, and a ferromagnetic material, respectively) [16,17], various types of structures exploiting manipulation of Abrikosov vortices in superconducting films [18,19] or magnetic flux trapped in a superconducting ring [20], and resistive states [21]. Several proposals of cryogenic memory consider integration of ferromagnetic materials and superconducting devices, such as, e. g., Josephson junctions. In particular, magnetic Josephson junctions (MJJs) wherein the barrier is made either with one or two magnetic layers have been considered as promising candidates for the cryogenic memory [5,7-9,21-24].

In our earlier publication, we have shown that memory function can be realized in a multi-terminal device in which a pseudo-spin valve (PSV) is placed on top of a Josephson junction [25]. This design has an advantage that the Josephson junction can be a usual Nb/Al/$AlO_x$/Nb junction used in SFQ circuits, which makes the device readily suitable for integration into superconducting electronics. The device [25] was not optimized, so that the difference in the values of the maximum Josephson current, $I_c$, for the "0" and "1" logical states was low. On the other hand, some theoretical [26] and experimental [27] works suggest that such a difference can be made very large, on the order of $I_c$, if, in one of the states, the minimum is obtained in the dependence of the critical current vs externally applied magnetic field, $H$, at $H=0$. Here, using a simplified device structure SISF, we show that this is indeed possible; furthermore, switching between the states with the maximum and minimum in the $I_c(H)$ dependence at about $H$=0 can be realized using the electric current control.

The devices were fabricated from Nb(75)Al/AlO$_x$(9)/Al(3)/Nb(68)/Py(10) multilayer structures sputter-deposited *in situ* onto oxidized Si substrates in a high-vacuum system; here, the numbers in parentheses denote the thicknesses of the respective layers in nm. The Al/AlO$_x$ tunnel barrier was formed by thermal oxidation of the Al overlayer whose thickness prior to oxidation was 9 nm. The devices were patterned using optical lithography, reactive ion etching of Nb, ion milling of permalloy (Py) and Al/AlO$_x$/Al, anodization and deposition of SiO$_2$ for proper insulation. Here we consider two sets of devices fabricated from the multilayers deposited in the same run but situated in different positions on the wheel holding the substrates and patterned in slightly different geometry. In addition, we also fabricated the control Nb(75)Al/AlO$_x$(9)/Al(3)/Nb(68) devices without any Py for comparison. In this case, the control line and the junction were separated with an Al/AlO$_x$/Al layer of the same thickness (10 nm) as the Py layer in the SISF devices. All devices were measured in a liquid He bath at 4.2 K.

Fig. 1(a) shows device schematic and its biasing. Nominal lateral dimensions of the Josephson junction are $L$=4 μm and $w$=2 μm. Normally, the area of the Py layer should be 2 μm × 2 μm, so that it should occupy one half of the Josephson junction area, and be situated at the end of the Josephson junction, with one edge stretching through the middle of the junction. In the devices of the 1$^{st}$ type which we consider next, the length of the Py layer is slightly shorter than 2 μm; because of this, one of its edges is slightly off the central axis of the Josephson junction. Fig. 1(b) shows an optical image of one of such devices (top view). The blue area highlights an "open" part of the Josephson junction, whereas the pink area corresponds to the position of the Py layer on top of the other end of the junction. Six devices of this type were characterized.

Fig. 2(a) shows typical *I-V* curve of the Josephson junction in the initial state (no magnetic field was applied). Current $I$ was applied and voltage $V$ was measured as shown in Fig. 1(a). We investigate how the dependence of the critical Josephson current, $I_c$, measured as a function of an external magnetic field, $H$, applied in the plane of the structure as shown in Fig. 1(a), is modified when the transport current, $I_{tr}$, is fed through the wire stretching across the junction length (electrode 2 in Fig. 1(b)). A family of $I_c(H)$ dependences measured for one of the devices is shown in Fig. 2(b) for different values of $I_{tr}$: $I_{tr}$=0 (black solid circles and red solid triangles), +10 mA (blue solid squares and magenta solid diamonds), +20 mA (olive solid triangles and violet stars), $I_{tr}$ = 0 after +25 mA was applied (black and red open circles), and $I_{tr}$=0 after $I_{tr}$ = -15 mA was applied (navy and orange open triangles). One can see the following main features: (*i*) suppression of $I_c$ with $I_{tr}$; (*ii*) broadening of the main lobe of the $I_c(H)$ dependence with $I_{tr}$, which develops in a minimum at about $H$=0 for two devices (see a family of $I_c(H)$ dependences for one of such devices in Fig. 2(c). Curves represented by olive and violet solid triangles for $I_{tr}$=+15 mA clearly display the minimum; and (*iii*) memory property. The latter (observed for four devices out of six measured) is manifested as a reduced $I_c$ after application of $I_{tr}$ = +25 mA (black and red open circles in Fig. 2(b)) and a restored $I_c$ close to its initial value (navy and orange open triangles) after application of $I_{tr}$ = -15 mA; sometimes, as in the particular case of Fig. 2(b), the restored $I_c$ is even slightly higher than the initial $I_c$ represented by black solid circles and red solid triangles.

The behavior of the control devices with the same size and the critical current density, $j_c$, in the same range (4.1 – 4.8 kA/cm$^2$) but without any Py layer is substantially different; see characteristics of a representative device of this kind in Fig. 2(d). The influence of $I_{tr}$ on the $I_c$ is very small compared with that for the devices involving Py; neither memory effect nor minimum in the $I_c(H)$ dependences for all three such devices was observed.

The above results clearly demonstrate the role of the magnetic layer in control of $I_c$ of the adjacent Josephson junction. The magnetic field of the control current $I_{tr}$ influences the orientation of the magnetic moment of the Py layer which can be aligned in the direction of the longer dimension of the Josephson junction; the arising fringe field acting perpendicular to the structure plane leads to suppression of the Josephson current and its distribution in such a way that the minimum in the $I_c(H)$ dependence may occur at about $H=0$, according to the earlier theoretical [26] and experimental [27] findings.

The occurrence of the minimum was more clearly observed for the second type of our devices (see an optical image of one of such devices in Fig. 3(a)). Note that in these devices the Py layer occupies approximately one half of the SIS junction area, so that one edge of the Py layer runs almost through the middle of the junction. Fig. 3(b) shows typical I-V curve of the device; Fig. 3(c) represents a result of application of $I_{tr}$ for this type of devices.

Black curve 1 in Fig. 3(c) is the initial $I_c(H)$ dependence before any current $I_{tr}$ was applied. Then we applied DC current $I_{tr} = +5$ mA ("+" corresponds to the current direction shown in Fig. 1(a)) and turned it off to 0. After that, we measured $I_c(H)$ dependence plotted as the red curve 2 in Fig. 3(c). Obviously, the $I_c$ modulation pattern experienced dramatic change, displaying the minimum near $H=0$. Next, we applied current $I_{tr} = -5$ mA, reduced it to 0, and measured the $I_c(H)$ dependence again; as a result, we obtained the blue curve 5 in Fig. 3(c). This curve is slightly shifted in $H$ as compared with the position of the original black curve, but the maximum in the main lobe is restored. We repeated the whole measurement cycle two more times and obtained magenta and orange curves 3 and 4 after application of $I_{tr} = +5$ mA, and the same blue curve 5 after subsequent application of $I_{tr} = -5$ mA. As one can see, although there is some dispersion of the $I_c(H)$ dependencies obtained after repeated application of "positive" $I_{tr}$, the minimum near $H=0$ is reproducible, which is the main result of this experiment. Also, the blue curve 5 obtained in each measurement cycle is reproducible. Note a large difference in $I_c$ at $H=0$ for the two states – one with the unsuppressed $I_c$ and the other with the suppressed $I_c$, which can reach $0.8I_c$ for the particular experimental case. This is important for building a memory based on the cell.

The physical origin of the observed behavior, as we understand it at present, is the effect of the fringe field [25,28] acting not only towards the suppression, but also on the distribution of the supercurrent in the Josephson junction. Specifically, in the case of the devices of the second kind, the current $I_{tr} = 5$ mA through the wire of about 5 μm wide creates magnetic field of 0.63 mT immediately below it and directed along the length $L$ of the Josephson junction. The field partially aligns the magnetization $M$ of the Py layer in the same direction. Although we do not know exactly magnetic properties of the Py layer, we can suggest, based on the literature data [29], that the relative permeability of this layer, $\mu$, at low magnetic fields is in the range of 100. Using the relation $M=\chi H$, where $\mu=1+\chi$, we estimate $M$ to be on the order of 60 mT. The fringe field $B_z$, perpendicular to the S layers, can be estimated as $B_z=2Md_F z/(x^2+z^2)$, where $d_F$ is the thickness of the F layer, $z$ is measured from the midpoint of F, $x$ is the distance from the edge of the F layer in the plane of S [28]. For $x=0$ and $z=d_F/2$, we obtain $B_z=240$ mT, which is well above the value of perpendicular $\mu_0 H_{c1}$ for the Nb film [25]. Therefore, this field can create a magnetic vortex (or vortices) penetrating one or both S layers of the Josephson junction, which, according to [26,27], can result in the minimum at $H=0$ in the $I_c(H)$ dependence.

Remarkably, application of "negative" $I_{tr}$ of the same (or even smaller, as in the case of the devices of the first type) magnitude than "positive" $I_{tr}$ can almost restore the initial shape of $I_c(H)$ dependence, which provides evidence that switching between the two states of the junction, one with the maximum and second with the minimum of $I_c$ at $H \approx 0$, can, potentially, be electrically controlled, which is important for possible application the device as a memory cell. Another important feature is that switching between the maximum and minimum provides a large discrimination between the two logical states, which will not be much affected by scaling down the device size.

In the current device embodiment, the $I_c(H)$ dependencies are not perfectly reproducible under influence of the current $I_{tr}$. We suggest the reason for this is the fact that we are dealing with devices of an intermediate size. In the Py island used in this work, several magnetic domains may be present [30]. In such a case, the domain structure in the samples having nominally the same dimensions may be different; each individual domain may considerably influence the total magnetic ordering, so that reproducible control of magnetic properties may be difficult. In order to have well reproducible properties, the device size has to be either considerably larger than the domain size, or smaller than that (i.e., below 1 μm in our case [30]). We believe this can be realized in the optimized devices.

In conclusion, we have fabricated and tested SISF type devices with memory properties, wherein a ferromagnetic (permalloy) island is positioned on top and at the edge of the SIS Josephson junction. We have shown that manipulation of the magnetization of the Py island with the help of the electric current passing through a control wire can lead to switching the main lobe of the modulation pattern of the Josephson critical current in an externally applied magnetic field between the shapes with the maximum and minimum in the vicinity of $H = 0$. This possibility of electric control, and a large discrimination between the Josephson critical current levels for the two states which will be preserved upon reduction of the device dimensions, is important for use in the memory cell. We suggest that reproducible device performance can be obtained in smaller devices where the dimensions of the magnetic island are on the order of or smaller than the magnetic domain size. Therefore, we believe that the optimized device is promising for building a compact cryogenic memory based on it.

## Acknowledgement

This research received support from NSF Grant DMR 1905742 and from the NSF DISCoVER Expedition award under Grant CCF-2124453. I. P. N. acknowledges the use of facilities of the Materials Research Center at Northwestern University, supervised by J. B. Ketterson and supported by NSF.

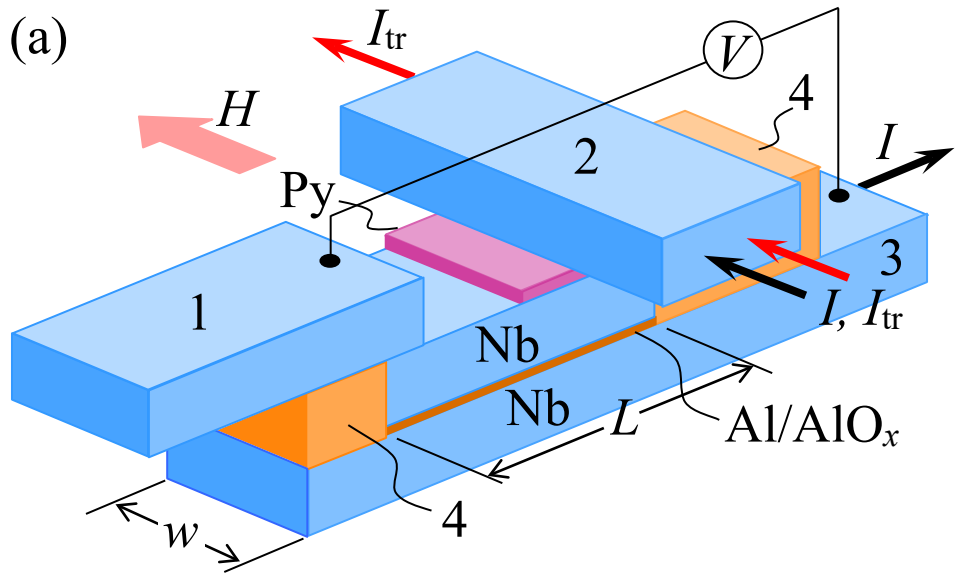

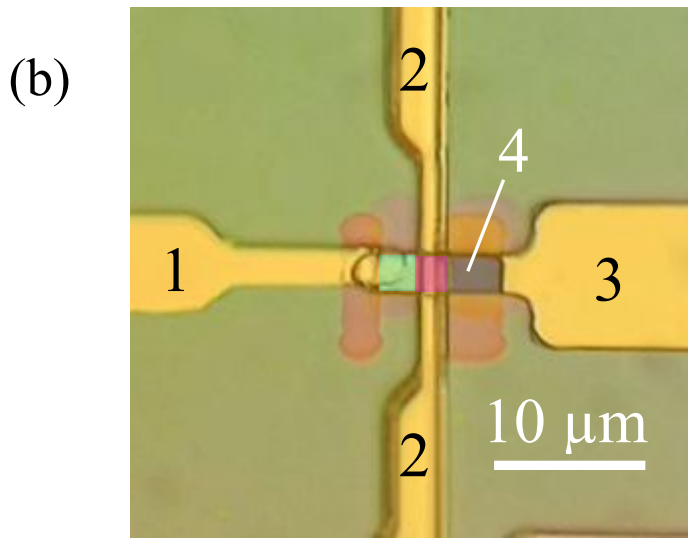

Fig. 1. (a) Schematic of the device and its biasing. Part numbers correspond to those in (b). (b) Optical image of an actual device (top view). Wires 1 and 2 are connected to the top electrode of the Josephson junction (through the Py layer in case of wire 2); wire 3 is connected to the bottom electrode of the Josephson junction; region 4 is a part of the bottom electrode covered with an insulator. The blue area highlights an "open" part of the Josephson junction, whereas the pink area corresponds to the position of the Py layer on top of the other end of the Josephson junction.

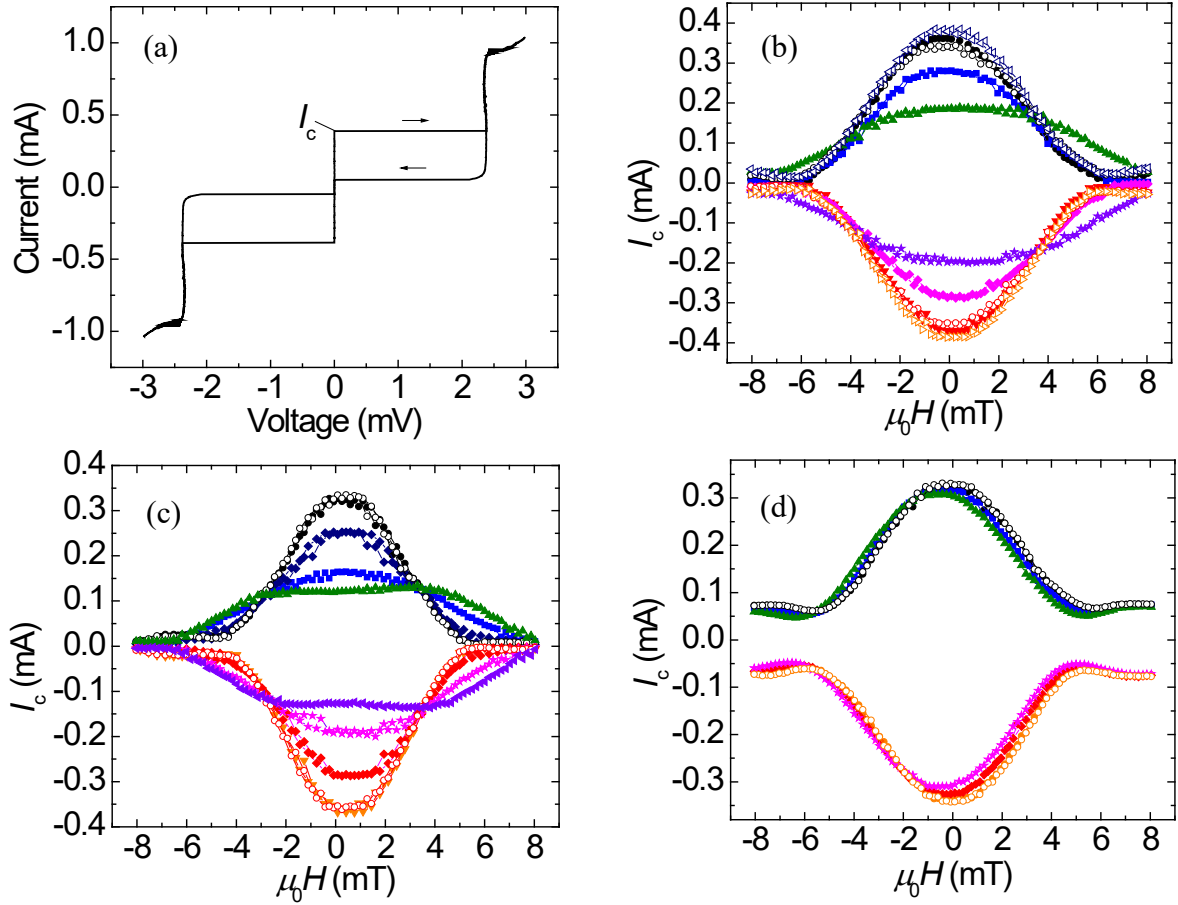

Fig. 2. (a) *I-V* curve of an NbAl/AlO$_x$/Al/Nb/Py device at 4.2 K. (b) $I_c(H)$ dependences of the same device in the initial state (black solid circles and red solid triangles) and when different values of $I_{tr}$ are applied: +10 mA (blue squares and magenta solid diamonds), +20 mA (olive solid triangles and violet stars), $I_{tr}$=0 after +25 mA was applied (black and red open circles), and after application of $I_{tr}$ = -15 mA (navy and orange open triangles). (c) $I_c(H)$ dependences of another device in the initial state at $I_{tr}$ =0 (black solid circles and orange solid triangles); +5 mA (navy and red solid diamonds); +10 mA (blue solid squares and magenta stars); +15 mA (green and violet solid triangles); and at $I_{tr}$ =0 after application of +25 mA (black and orange open circles). Note the minimum at about H=0 in the curves for $I_{tr}$ =+15 mA. (d) $I_c(H)$ dependences of a device without any Py layer for $I_{tr}$ =0 (black and orange solid circles); +10 mA (blue solid squares and red solid diamonds); +20 mA (olive solid triangles and magenta stars), and $I_{tr}$ =0 after application of +25 mA (black and orange open circles).

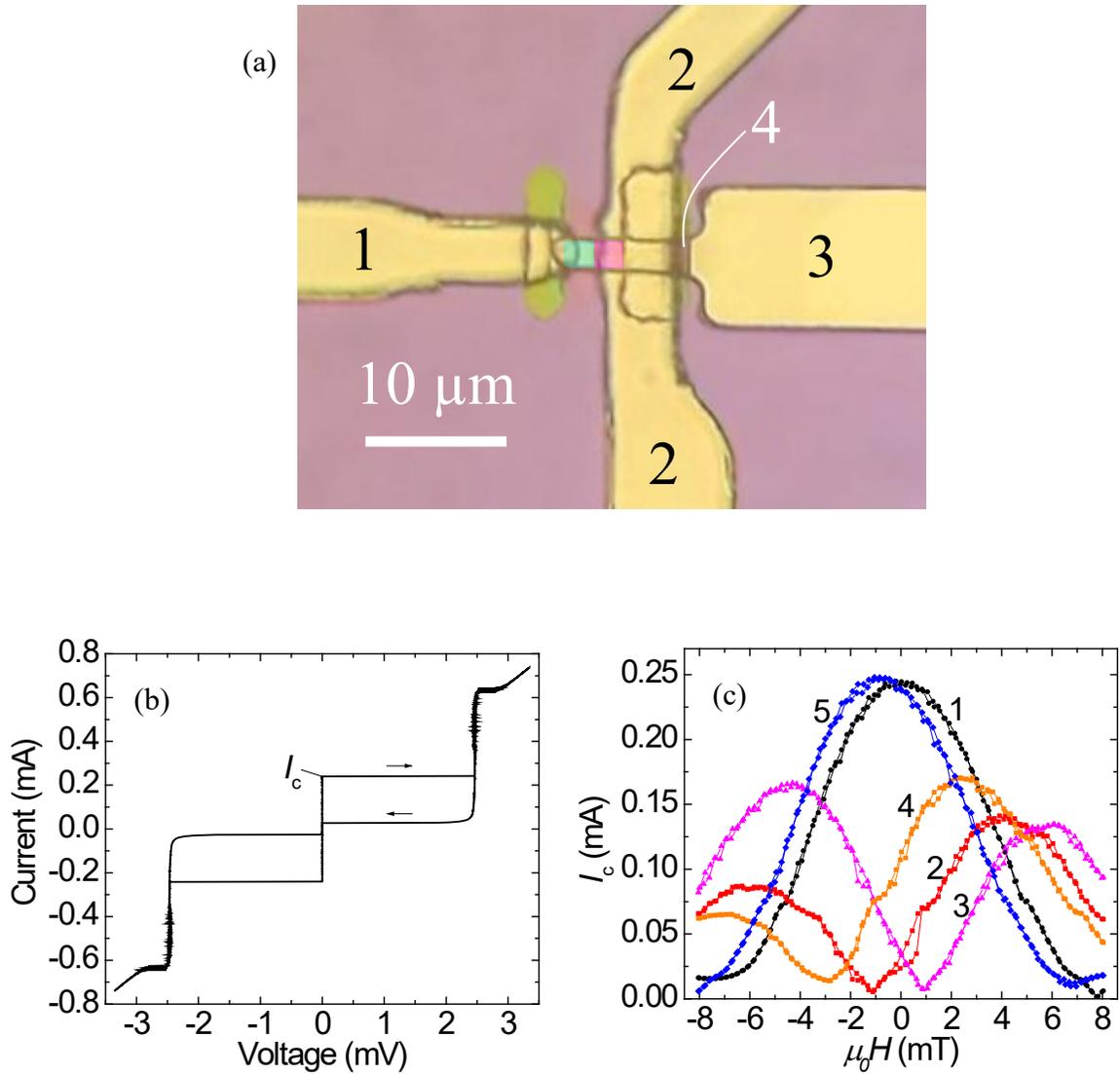

Fig. 3. (a) Optical image of an actual SISF device of the second type (top view). The blue area highlights an "open" part of the Josephson junction, whereas the pink area corresponds to the position of the Py layer on top of the other end of the Josephson junction. (b) *I-V* curve of an NbAl/AlO$_x$/Al/Nb/Py junction at 4.2 K. (c) $I_c(H)$ dependences of the same device in the initial state (black curve 1); after application of $I_{tr}$ = + 5 mA (red, magenta, and orange curves 2 – 4, respectively); and after application of $I_{tr}$ = - 5 mA (blue curve 5). See main text for details.